\documentclass[prd,twocolumn,showpacs,preprintnumbers,superscriptaddress,nofootinbib,floatfix]{revtex4-1}

\usepackage{amssymb,amsmath,verbatim,mathtools,needspace,enumitem,etoolbox,graphicx,physics,microtype,afterpage}

\usepackage[dvipsnames, usenames]{xcolor}

\definecolor{linkcolor}{rgb}{0.0,0.3,0.5}
\usepackage[unicode, colorlinks=true, linkcolor=linkcolor, citecolor=linkcolor, filecolor=linkcolor,urlcolor=linkcolor, pdfusetitle, linktocpage, breaklinks]{hyperref}
\usepackage[all]{hypcap}
\usepackage[T1]{fontenc}
\usepackage[utf8]{inputenc}
\usepackage{tabularx}
\allowdisplaybreaks
\usepackage[english]{babel}
\usepackage{eufrak}
\usepackage{caption}
\usepackage{subcaption}
\usepackage{aas_macros}
\def\square{\mathchoice\sqr54\sqr54\sqr{2.1}3\sqr{1.5}3}
\def\sqr#1#2{{\vcenter{\vbox{\hrule height.#2pt\hbox{\vrule
width.#2pt height#1pt \kern#1pt\vrule width.#2pt}\hrule height.#2pt}}}}
\def\square{\mathchoice\sqr54\sqr54\sqr{2.1}3\sqr{1.5}3}

\makeatletter

\@addtoreset{equation}{section}
\makeatother

\begin{document}

\title{d+1 formalism in Einstein-scalar-Gauss-Bonnet gravity}
\author{F\'elix-Louis Juli\'e}
\email{fjulie@jhu.edu}
\author{Emanuele Berti}
\email{berti@jhu.edu}
\affiliation{Department of Physics and Astronomy, Johns Hopkins University, 3400 N. Charles Street, Baltimore, MD 21218, USA}

\date{\today}

\begin{abstract}
We present the $d+1$ formulation of Einstein-scalar-Gauss-Bonnet (ESGB) theories in dimension $D=d+1$ and for arbitrary (spacelike or timelike) slicings. We first build an action which generalizes those of Gibbons-Hawking-York and Myers to ESGB theories, showing that they can be described by a Dirichlet variational principle. We then generalize the Arnowitt-Deser-Misner (ADM) Lagrangian and Hamiltonian to ESGB theories, as well as the resulting $d+1$ decomposition of the equations of motion. Unlike general relativity, the canonical momenta of ESGB theories are nonlinear in the extrinsic curvature. This has two main implications: (i) the ADM Hamiltonian is generically multivalued, and the associated Hamiltonian evolution is not predictable; (ii) the ``$d+1$'' equations of motion are quasilinear, and they may break down in strongly curved, highly dynamical regimes.
Our results should be useful to guide future developments of numerical relativity for ESGB gravity in the nonperturbative regime.
\end{abstract}

\maketitle

\section{Introduction}

The study of modifications of general relativity is well motivated by some of the most outstanding puzzles in theoretical physics (where such modifications are often invoked e.g. in the quest to quantize gravity, or to solve the information paradox) and in observational cosmology (where the nature of dark matter and dark energy is unclear, although both of them seem to interact only gravitationally)~\cite{Joyce:2014kja,Berti:2015itd,Koyama:2015vza}.

The detection of gravitational waves by the LIGO/Virgo collaboration finally allows us to test observational signatures of modified gravity in the dynamical, strong-field regime of a coalescing compact binary system. Given our limited understanding of the state of matter in neutron stars, the cleanest tests involve gravitational wave observations of black-hole binary mergers (which, moreover, are the vast majority of detected events so far) with Earth-based and future space-based interferometers~\cite{Yunes:2013dva,Yunes:2016jcc,Gair:2012nm,Berti:2018cxi,Berti:2018vdi,Berti:2019xgr,Barausse:2020rsu}. Most tests of general relativity obtained so far are essentially ``null tests'': they place bounds on phenomenological parameters that would be different from zero (or unity) if general relativity were not correct~\cite{TheLIGOScientific:2016src,LIGOScientific:2019fpa}.

Going beyond this sort of null tests requires the calculation of gravitational waveforms for specific modified theories of gravity. In particular, the numerical simulation of the field equations is necessary to take into account the full nonlinear dynamics of the merger.
A preliminary step in this direction is the $d+1$ decomposition of the field equations and the study of the corresponding Cauchy problem (in this paper we will focus on a $d+1$ dimensional theory, but the $d=3$ case is the most interesting from a phenomenological point of view).

The Cauchy problem is known to be well-posed for a very limited class of theories whose metric sector is the same as in general relativity (see Refs.~\cite{MSM_1927__25__1_0,foures-bruhat1952,foures-bruhat1956} for the earliest $3+1$ decomposition of Einstein's equations and the study of their Cauchy problem). These include the simplest scalar-tensor (ST) theories of gravity~\cite{Salgado:2008xh,Shibata:2013pra} and Einstein-Maxwell-dilaton (EMD) theories~\cite{Jai-akson:2017ldo,Hirschmann:2017psw}. However there are ``no-hair theorems'' which imply that black-hole binary inspirals in ST theories must reduce to general relativity, unless one enforces the presence of ``ad hoc'' scalar field clouds just before merger~\cite{Healy:2011ef} or nontrivial boundary conditions~\cite{Horbatsch:2011ye,Berti:2013gfa}. Black-hole binary mergers in EMD theories were evolved numerically in Ref~\cite{Hirschmann:2017psw} and found to be nearly indistinguishable from their general relativistic counterparts for small values of the electric charge, but their scalar cosmological environment may play a crucial role~\cite{Julie:2017rpw}.

In this paper we study the $d+1$ formulation of a class of theories whose metric sector differs from that of general relativity: Einstein-scalar-Gauss-Bonnet (ESGB) gravity.  These theories supplement the Einstein-Hilbert action with one single scalar degree of freedom $\varphi$ coupled to the Gauss-Bonnet scalar $\mathcal R_{\rm GB}^2=R^{\mu\nu\rho\sigma}R_{\mu\nu\rho\sigma}-4R^{\mu\nu}R_{\mu\nu}+R^2$ through a coupling of the form $f(\varphi)\mathcal R_{\rm GB}^2$. They pass Solar System tests~\cite{Sotiriou:2006pq} as well as the stringent gravitational-wave propagation tests from GW170817~\cite{Abbott:2018lct} (see e.g.~\cite{Franchini:2019npi,Noller:2019chl}), and they are interesting for various reasons.

The theory above with an exponential coupling function (also known as Einstein-dilaton-Gauss-Bonnet gravity) corresponds to the bosonic sector of heterotic string theory~\cite{Gross:1986mw}. Black hole solutions in this theory have long been known to differ from GR~\cite{Mignemi:1992nt,Kanti:1995vq} (see also~\cite{Antoniou:2017acq}). Theories with a generic coupling function $f(\varphi)$ have recently attracted interest because they can exhibit scalarization in vacuum: black hole solutions can reduce to those of general relativity in certain regions of parameter space, and spontaneously scalarize to very different solutions in others~\cite{Silva:2017uqg,Doneva:2017bvd,Cunha:2019dwb,Collodel:2019kkx}. Studies of the radial stability of these black hole solutions led to two interesting findings: (i) the stability depends crucially on the choice of the coupling function, and (ii) the hyperbolicity of the equations of motion of the perturbations seems to be broken when the coupling is large~\cite{Blazquez-Salcedo:2018jnn,Silva:2018qhn,Macedo:2019sem}. This hints at the possibility that the well-posedness of the field equations may depend on the strength of the coupling to the Gauss-Bonnet scalar.
This is crucial, because most analytical~\cite{Yagi:2011xp,Yagi:2015oca} and numerical~\cite{Witek:2018dmd,Okounkova:2019zep,Okounkova:2019zjf,Okounkova:2020rqw} studies of black-hole binaries in ESGB 
have relied on a weak coupling expansion (but see~\cite{Julie:2019sab} for a post-Newtonian calculation valid in principle for all couplings, and Refs.~\cite{Cayuso:2017iqc,Allwright:2018rut,Bernard:2019fjb} for different attempts to find global solutions and control higher-order gradients in modified theories of gravity).

Recent work studied the hyperbolicity of ESGB gravity under specific symmetries~\cite{Ripley:2019hxt,Ripley:2019irj,Ripley:2019aqj} as well as its well-posed formulation in the small-coupling regime~\cite{Papallo:2017qvl,Kovacs:2020pns,Kovacs:2020ywu}.
Our goal in this paper is to go beyond these approximations by developing an extension of the ADM formalism~\cite{Arnowitt:1959ah,Arnowitt:1962hi} for ESGB gravity and to provide, for the first time, their $``d+1"$ field equations.

The plan of the paper is as follows. In Sec.~\ref{sectionDirichlet} we extend the actions of Gibbons-Hawking-York \cite{Gibbons:1976ue,York:1972sj} and Myers \cite{Myers:1987yn} to ESGB gravity, and we formulate a Dirichlet variational principle. In Sec.~\ref{sectionADMformalism} we develop the ADM formalism for ESGB gravity, and in Sec.~\ref{section3+1eom} we write down the $``d+1"$ equations of motion. One of our main results is that the canonical momenta in ESGB gravity are nonlinear in the extrinsic curvature. As a consequence the Hamiltonian is multivalued, and the field equations become quasilinear. In Sec.~\ref{sec:conclusions} we comment and speculate on the implications of these results, and we discuss possible directions for future work.

\section{Variational principle and boundary terms\label{sectionDirichlet}}

In vacuum, ESGB theories are described by the action
\begin{equation}
  \!I=\!\int_{\mathcal M}\!\!\!\!\frac{d^Dx\sqrt{-g}}{16\pi}\Big(R-2g^{\mu\nu}\partial_\mu\varphi\partial_\nu\varphi+\alpha f(\varphi) \mathcal R^2_{\rm GB}\Big)\, ,
  \label{bulkAction}
\end{equation}
where we set $G= c= 1$. In standard notation, $R$ is the Ricci scalar on the $D$-dimensional manifold $\mathcal M$ with metric $g_{\mu\nu}$, inverse metric $g^{\mu\nu}$ and metric determinant $g$, and $\mathcal R_{\rm GB}^2=R^{\mu\nu\rho\sigma}R_{\mu\nu\rho\sigma}-4R^{\mu\nu}R_{\mu\nu}+R^2$ denotes the Gauss-Bonnet scalar, $R^\mu_{\ \nu\rho\sigma}$ being the Riemann tensor. The fundamental constant $\alpha$ (assumed to be positive without loss of generality) has the dimensions of a length squared, and $f(\varphi)$ is a dimensionless function defining the theory.

In the following, it will be useful to rewrite the Gauss-Bonnet scalar as
\begin{equation}
\mathcal R_{\rm GB}^2=R^{\mu\nu\rho\sigma}P_{\mu\nu\rho\sigma}\,,\label{GBusingP}
\end{equation}
 where
\begin{align}
  P^{\mu\nu}_{\ \ \,\,\rho\sigma}&=R^{\mu\nu}_{\ \ \,\rho\sigma}-2\delta^\mu_{[\rho}R_{\sigma]}^\nu+2\delta^\nu_{[\rho}R_{\sigma]}^\mu+\delta^\mu_{[\rho}\delta_{\sigma]}^\nu R\nonumber\\
&=\frac{1}{4}\delta^{\mu\nu\alpha_1\alpha_2}_{\rho\sigma\beta_1\beta_2}R_{\quad\ \,\alpha_1\alpha_2}^{\beta_1\beta_2}\ .
\end{align}
Here brackets denote antisymmetrization, as in $\delta^\mu_{[\rho}\delta_{\sigma]}^\nu=\frac{1}{2}(\delta^\mu_{\rho}\delta_{\sigma}^\nu-\delta^\mu_{\sigma}\delta_{\rho}^\nu)$, and $\delta^{\alpha_1\cdots\alpha_n}_{\beta_1\cdots\beta_n}=n!\,\delta^{\alpha_1}_{[\beta_1}\cdots\delta^{\alpha_n}_{\beta_n]}$ is the generalized Kronecker symbol, i.e. the determinant of the $n\times n$ matrix $M$ built from ordinary Kronecker ``deltas'', with elements $M^i_j=\delta^{\alpha_i}_{\beta_j}$, which is antisymmetric under exchange of its upper (and lower) indices.
The quantity $P_{\mu\nu\rho\sigma}$ has the symmetries of the Riemann tensor and it is divergence-free: denoting by $\nabla_\mu$ the covariant derivative associated to $g_{\mu\nu}$, the Bianchi identities imply $\nabla_\mu P^\mu_{\ \, \nu\rho\sigma}=0$.

The variation of the action (\ref{bulkAction}) with respect to $g^{\mu\nu}$ reads
\begin{align}
\delta_{(g)}I=\frac{1}{16\pi}\int_{\mathcal M}\!\!\!d^Dx\sqrt{-g}\left(E_{\mu\nu}\delta g^{\mu\nu}+\nabla_\mu V^\mu\right)\,,\label{covariantVariation}
\end{align}
where
\begin{align}
\quad E_{\mu\nu}&=G_{\mu\nu}-2\partial_\mu\varphi\partial_\nu\varphi+g_{\mu\nu}(\partial\varphi)^2\nonumber\\
&+\alpha\left(f(\varphi)H_{\mu\nu}+4P_{\mu\alpha\nu\beta}\nabla^\alpha\nabla^\beta f(\varphi)\right)\,,\label{einsteinFieldEqn}
\end{align}
$G_{\mu\nu}$ denotes the Einstein tensor and 
\begin{align}
H^\mu_\nu&= 2R^\mu_{\ \alpha\beta\gamma}P_{\nu}^{\ \alpha\beta\gamma}-\frac{1}{2}\delta^\mu_\nu\mathcal R_{\rm GB}^2\nonumber\\
&=-\frac{1}{8}\delta^{\mu\,\alpha_1\alpha_2\alpha_3\alpha_4}_{\,\nu\,\beta_1\beta_2\beta_3\beta_4}R_{\quad\ \,\alpha_1\alpha_2}^{\beta_1\beta_2}R_{\quad\ \,\alpha_3\alpha_4}^{\beta_3\beta_4}\label{lanczosTensor}
\end{align}
is the divergence-free~\cite{Lovelock:1971yv} Lanczos tensor, which vanishes identically in dimension $D\leqslant 4$, as obvious from its expression above in terms of the rank-five generalized Kronecker symbol.
Equation~(\ref{covariantVariation}) follows from the identity $\delta R^\mu_{\ \nu\rho\sigma}=2\nabla_{[\rho}\delta\Gamma^\mu_{\sigma]\nu}$ with $\delta\Gamma^\mu_{\nu\rho}=\frac{1}{2}g^{\mu\lambda}(\nabla_\nu\delta g_{\lambda\rho}+\nabla_\rho\delta g_{\lambda\nu}-\nabla_\lambda\delta g_{\nu\rho})$, integration by parts and the properties of $P_{\mu\nu\rho\sigma}$.

As far as we know, the second term in the integrand on the right hand side of Eq.~(\ref{covariantVariation}) was not previously considered in the ESGB literature. It is the divergence of the four-vector
\begin{align}
V^\mu&=\left[g^{\mu\alpha}g^{\lambda\beta}-g^{\mu\lambda}g^{\alpha\beta}-4\alpha f(\varphi)P^{\lambda\alpha\mu\beta}\right]\nabla_\lambda\delta g_{\alpha\beta}\nonumber\\
&+\left[4\alpha P^{\mu\alpha\lambda\beta}\nabla_\lambda f(\varphi)\right]\delta g_{\alpha\beta}\ ,
\end{align}
and can therefore be evaluated on the $d=D-1$ dimensional boundary $\partial\mathcal M$ of $\mathcal M$.
Let us choose for simplicity an adapted gaussian coordinate system $x^\mu=\{w,x^i\}$ such that $w$ is constant on  $\partial\mathcal M$:
\begin{equation}
ds^2=\epsilon N^2dw^2+h_{ij}dx^idx^j\,,\label{gaussianCoord}
\end{equation}
where $\epsilon= 1$ if $\partial \mathcal M$ is timelike, $\epsilon=-1$ if $\partial \mathcal M$ is spacelike, and such that $\sqrt{-g}=N\sqrt{|h|}$. Then $P^{\mu\nu\rho\sigma}$ can be decomposed using the Gauss-Codazzi-Mainardi identities \cite{Gourgoulhon:1453298,Deruelle:2018ltn}
\begin{subequations}
\begin{align}
R_{ijkl}&=\bar R_{ijkl}-\epsilon\left(K_{ik}K_{jl}-K_{il}K_{jk}\right)\ ,\label{gaussCodazziMainardi1}\\
R_{ijkw} &=N\!\left(\bar\nabla_i K_{jk}-\bar\nabla_j K_{ik}\right)\ ,\\
R_{iw jw} &= -N \partial_wK_{ij}+N^2K_{ik}K^k_j-\epsilon N\bar\nabla_i\bar\nabla_j N\ ,\label{gaussCodazziMainardi3}
\end{align}\label{gaussCodazziMainardi}%
\end{subequations}
where from now on latin indices are lowered with the induced metric $h_{ij}$ (with inverse $h^{ij}$ and determinant $h$), bars denote intrinsic quantities built out of $h_{ij}$ (as in $\bar\nabla_i W^j=\partial_iW^j+\bar\Gamma_{ik}^jW^k$), and $K^{ij}=- \frac{1}{2N}\partial_w h^{ij}$ is the extrinsic curvature.
Introducing by analogy the notation $K_{\varphi}=-\frac{1}{2N}\partial_w\varphi$, the variation (\ref{covariantVariation}) of the ESGB action with respect to $g^{\mu\nu}$ yields
\begin{align}
\delta_{(g)}I&=\frac{1}{16\pi}\int_{\mathcal M}\!\!\!d^Dx\sqrt{-g}E_{\mu\nu}\delta g^{\mu\nu}\label{adaptedVariation}\\*
&+\frac{1}{16\pi}\int_{\mathcal \partial M}\!\!\!\!\!d^d x\left[\sqrt{|h|}\pi_{ij}\delta h^{ij}-\delta_{(g)}\left(\!\sqrt{|h|}\,Q\right)\right]\, ,\nonumber
\end{align}
where
\begin{widetext}
\begin{subequations}
\begin{align}
\epsilon Q&=2K+2\alpha f(\varphi)\,\delta^{i_1i_2i_3}_{j_1j_2j_3}K^{j_1}_{i_1}\left(\bar R^{j_2j_3}_{\ \ \ \ i_2i_3}-\frac{2\epsilon}{3}K^{j_2}_{i_2}K^{j_3}_{i_3}\right)\ ,\label{boundaryTermCovariant}\\
\epsilon\pi^i_j&=-\delta^{i\,i_1}_{j\,j_1}K^{j_1}_{i_1}+2\alpha\,\delta^{i\,i_1 i_2}_{j\,j_1j_2}\left[2K^{j_1}_{i_1}\bar\nabla^{j_2}\bar\nabla_{\!i_2}f(\varphi)+f'(\varphi)\,K_\varphi\left(\bar R^{j_1 j_2}_{\ \ \ \ i_1 i_2}-2\epsilon K^{j_1}_{i_1}K^{j_2}_{i_2}\right)\right]\nonumber\\
&\hspace*{2cm}-\alpha f(\varphi)\, \delta^{i\,i_1i_2i_3}_{j\,j_1j_2j_3}K^{i_1}_{j_1}\left(\bar R^{j_2 j_3}_{\ \ \ \ i_2 i_3}-\frac{2\epsilon}{3}K^{j_2}_{i_2}K^{j_3}_{i_3}\right)\ ,\label{momentumMetric}
\end{align}\label{momenta}%
\end{subequations}
\end{widetext}
and where we defined $K= h^{ij}K_{ij}$ and $f'(\varphi)= df/d\varphi$. Note that in the boundary term of Eq.~(\ref{adaptedVariation}) we have ignored the divergence of a $d$-vector $\bar\nabla_i W^i$, because its integral over the closed boundary $\partial\mathcal M$ vanishes.

The results above reduce to those of Einstein-Gauss-Bonnet gravity when $f(\varphi)$ is a constant \cite{Myers:1987yn,Davis:2002gn,Deruelle:2018vtt,guilleminot}. Note also that when $D\leqslant 4$ (i.e., $d\leqslant 3$) the second line of Eq.~(\ref{momentumMetric}) vanishes identically.
 
From (\ref{adaptedVariation}) and (\ref{boundaryTermCovariant}), we see that extremizing the action $I$ requires fixing both the metric $h^{ij}$ and its normal derivative $\partial_wh^{ij}=-2NK^{ij}$ on $\partial\mathcal M$, at odds with the Einstein field equations $E_{\mu\nu}=0$ being of second order only.

Let us now generalize Eq.~(\ref{bulkAction}) as follows:
 \begin{equation}
  I_{\rm ESGB}=I+\frac{1}{16\pi}\int_{\mathcal \partial M}\!\!\!\!\!d^d x\sqrt{|h|}\,Q\ ,\label{actionMyers}
 \end{equation}
with $Q$ given by Eq.~(\ref{boundaryTermCovariant}). This action extends those of Gibbons-Hawking-York \cite{Gibbons:1976ue,York:1972sj} and Myers \cite{Myers:1987yn} to ESGB gravity, and it allows to obtain the field equations by means of a \textit{Dirichlet variational principle}:
\begin{itemize}
\item[(i)] The variation of (\ref{actionMyers}) with respect to $g^{\mu\nu}$ reads
\begin{align}
\delta_{(g)}I_{\rm ESGB}&=\frac{1}{16\pi}\int_{\mathcal M}\!\!\!d^Dx\sqrt{-g}E_{\mu\nu}\delta g^{\mu\nu}\nonumber\\
&+\frac{1}{16\pi}\int_{\mathcal \partial M}\!\!\!\!d^d x\sqrt{|h|}\pi_{ij}\delta h^{ij}\ ,\label{variationMetric_Iesgb}
\end{align}
where $E_{\mu\nu}$ and $\pi_{ij}$ are given in Eqs.~(\ref{einsteinFieldEqn}) and (\ref{momentumMetric}). The action is extremal when the metric satisfies the second-order generalized Einstein field equations $E_{\mu\nu}=0$ together with Dirichlet boundary conditions, i.e. $\delta h^{ij}\vert_{\partial\mathcal M}=0$;

\item[(ii)] The variation of Eq.~(\ref{actionMyers}) with respect to $\varphi$ reads
\begin{align}
\delta_{(\varphi)}I_{\rm ESGB}&=\frac{1}{16\pi}\int_{\mathcal M}\!\!\!\!d^Dx\sqrt{-g}E_{\varphi}\delta \varphi\nonumber\\*
&+\frac{1}{16\pi}\int_{\mathcal \partial M}\!\!\!\!\!d^d x\sqrt{|h|}\pi_{\varphi}\delta \varphi\ ,\label{variationPHI}
\end{align}
where
\begin{subequations}
\begin{align}
\qquad\qquad E_\varphi&=4\,\Box\,\varphi+\alpha f'(\varphi)\mathcal R_{\rm GB}^2\ ,\label{eomKGcovariante}\\
\epsilon\pi_\varphi&=8K_\varphi+2\alpha  f'(\varphi)\delta^{i_1i_2i_3}_{j_1j_2j_3} K^{j_1}_{i_1}\nonumber\\
&\times\left(\bar R^{j_2j_3}_{\ \ \ \ i_2i_3}-\frac{2\epsilon}{3}K^{j_2}_{i_2}K^{j_3}_{i_3}\right)\ ,\label{momentumPhi}
\end{align}
\end{subequations}
and we defined (as usual) $\Box=\nabla^\mu\nabla_\mu$. The action is extremal when the scalar field satisfies the second-order generalized Klein-Gordon equation $E_{\varphi}=0$ together with Dirichlet boundary conditions, i.e. $\delta \varphi\vert_{\partial\mathcal M}=0$.
\end{itemize}

The action (\ref{actionMyers}) is the first new result of this paper. It shows that ESGB theories can be consistently described by a Dirichlet variational principle.

From the boundary terms in Eqs.~(\ref{variationMetric_Iesgb}) and (\ref{variationPHI}) we see that ESGB theories propagate one scalar degree of freedom and $d(d+1)/2$ metric degrees of freedom (the independent components of $h^{ij}$ in coordinates adapted to the boundary). As usual with covariant gravity theories, the $D=d+1$ remaining components of $g^{\mu\nu}$ can be fixed at will.

Note that the presence of the boundary term in Eq.~(\ref{actionMyers}) affects the quantities $\pi_{ij}$ and $\pi_\varphi$: for example, varying only the bulk action $I$ with respect to $\varphi$ would yield (\ref{variationPHI}) with $\epsilon\pi_\varphi=8K_\varphi$, instead of its expression (\ref{momentumPhi}) above.

\section{ADM formalism\label{sectionADMformalism}}

Consider first the Einstein-Hilbert action supplemented by the Gibbons-Hawking-York (GHY) boundary term:
\begin{equation}
I_{\rm GR}=\frac{1}{16\pi}\int_{\mathcal M}\!\!\!d^Dx\sqrt{-g}R+\frac{1}{16\pi}\int_{\mathcal \partial M}\!\!\!\!\!d^d x\sqrt{|h|}\,Q_{\rm GHY}\ ,\label{grGHYaction}
\end{equation}
with $Q_{\rm GHY}=2\epsilon K$, the limit of $Q$ given in Eq.~(\ref{boundaryTermCovariant}) above as $\alpha\to 0$. As recalled in the previous section, this boundary term ensures that extremizing the action (\ref{grGHYaction}) yields Einstein's field equations $G_{\mu\nu}=0$ when we impose Dirichlet boundary conditions.

In gaussian coordinates 
(\ref{gaussianCoord}), which foliate $\mathcal M$ with surfaces $\Sigma_w$ of constant $w$, we can use the Gauss-Codazzi-Mainardi identities (\ref{gaussCodazziMainardi}) to decompose the Einstein-Hilbert Lagrangian density on $\Sigma_w$ as
\begin{align}
\hspace*{-0.2cm} R=\bar R+\epsilon\,\delta^{i_1i_2}_{j_1j_2}K^{j_1}_{i_1}K^{j_2}_{i_2}-\frac{\partial_w(\sqrt{|h|}2\epsilon K)}{N\!\sqrt{|h|}}-\frac{2\bar\Box N}{N}\,.
\end{align}

Now define the closed boundary as the union $\partial \mathcal M=\Sigma_{w_i}\cup\Sigma_{w_f}\cup\mathcal B$ of the surfaces $w=w_i$ and $w=w_f$ and their complement $\mathcal B$ (which is a timelike cylinder when $w$ is a time coordinate). By plugging the expression above into Eq.~(\ref{grGHYaction}) with $\sqrt{-g}=N\sqrt{|h|}$ we find that $I_{\rm GR}$ coincides with the Arnowitt-Deser-Misner (ADM) bulk action~\cite{Arnowitt:1962hi}
\begin{equation}
I_{\rm GR}=\frac{1}{16\pi}\int_{\mathcal M}\!\!\!d^Dx\,N\!\sqrt{|h|}\left(\bar R+\epsilon\,\delta^{i_1i_2}_{j_1j_2}K^{j_1}_{i_1}K^{j_2}_{i_2}\right) ,\label{ADMactionGR}
\end{equation}
modulo a contribution on $\mathcal B$ which we can discard for our purposes (but which is essential to define the ADM mass~\cite{Poisson:2009pwt}). The integrand of Eq (\ref{ADMactionGR}) is the ADM Lagrangian, and it does not depend on the second-order $w$-derivatives of the metric.

Since the variation of $I_{\rm ESGB}$ also yields a Dirichlet variational principle, it must be possible to rewrite it in ADM-like form as:
\begin{align}
I_{\rm ESGB}=\frac{1}{16\pi}\int_{\mathcal M}\!\!\! d^Dx\sqrt{|h|}\, \mathcal L\label{actionADM}
\end{align}
(modulo boundary terms on $\mathcal B$), where $\mathcal L$ should depend at most on the fields' first-order normal derivatives $\partial_w h^{ij}$ and $\partial_w\varphi$. However, the ESGB Lagrangian density (\ref{bulkAction}) is quadratic in the Riemann tensor, and reproducing the decomposition above would involve cumbersome  calculations.
As we will show momentarily, it is straightforward to compute $\mathcal L$ if we follow instead the procedure recently developed in Ref.~\cite{guilleminot} in the context of Lovelock gravity.

Indeed, from Eqs.~(\ref{variationMetric_Iesgb}) and (\ref{variationPHI}) we first get that $\pi_{ij}$ and $\pi_\varphi$ are respectively the conjugate momentum densities to $h^{ij}$ and $\varphi$, as is well-known in classical mechanics~\cite{Landau1976Mechanics}. Therefore $\mathcal L$ must satisfy
\begin{subequations}
\begin{align}
\frac{\partial \mathcal L}{\partial(\partial_w h^{ij})}&=\pi_{ij}\ ,\\*
\frac{\partial \mathcal L}{\partial(\partial_w\varphi)}&=\pi_\varphi\ ,
\end{align}\label{eqDiffLagrangian}%
\end{subequations}
where $\partial_w h^{ij}=-2N\!K^{ij}$ and $\partial_w\varphi=-2N\!K_{\varphi}$ in the gaussian coordinates introduced above. The system~(\ref{eqDiffLagrangian}) is integrable: the quantities $\pi_{ij}$ and $\pi_\varphi$ given in (\ref{momentumMetric}) and (\ref{momentumPhi}) satisfy the identities
\begin{subequations}
\begin{align}
\frac{\partial\pi_{ij}}{\partial K_{ab}}&=\frac{\partial\pi^{ab}}{\partial K^{ij}}\ ,\\
\frac{\partial\pi_{ij}}{\partial K_\varphi}&=\frac{\partial\pi_\varphi}{\partial K^{ij}}\ .
\end{align} \label{commutationMomenta}%
\end{subequations}
Their explicit expressions are given in Eqs. (\ref{dPidK}) below.

Therefore we can integrate Eq.~(\ref{eqDiffLagrangian}) to get $\mathcal L$, modulo terms $\bar{\mathcal L}$ which must be identical to the part of the ESGB Lagrangian (\ref{bulkAction}) which, when calculated in gaussian coordinates, depends only on the intrinsic geometry of the constant-$w$ surfaces $\Sigma_w$:
\begin{align}
N^{-1}\bar{\mathcal L}&=\bar R-2\bar\nabla^k\varphi\bar\nabla_k\varphi+\frac{\alpha}{4} f(\varphi)\delta^{i_1i_2i_3i_4}_{j_1j_2j_3j_4}\bar R^{j_1j_2}_{\ \ \ \ i_1i_2}\bar R^{j_3j_4}_{\ \ \ \ i_3i_4}\nonumber\\
&-2\alpha\,\delta^{i_1i_2i_3}_{j_1j_2j_3}\bar R^{j_1j_2}_{\ \ \ \ i_1i_2}\bar\nabla^{j_3}\bar\nabla_{i_3}f(\varphi)\ ,
\end{align}
where the first three terms follow trivially from Eq.~(\ref{bulkAction}), while obtaining the fourth term requires introducing a nonconstant lapse $N$ and integrating the last term of the Gauss-Codazzi-Mainardi identity (\ref{gaussCodazziMainardi3}) by parts.

In the third and last step, we can generalize the result to arbitrary ADM metric variables \cite{FouresBruhat:1952zz,Arnowitt:1959ah}
\begin{equation}
ds^2=\epsilon N^2 dw^2+h_{ij}(dx^i+N^idw)(dx^j+N^jdw)\label{admCoordinates}
\end{equation}
through the redefinitions
\begin{subequations}
\begin{align}
K_{ij}&=\frac{1}{2N}\left(\partial_w-\mathcal L_{N^k}\right)h_{ij}\nonumber\\
&=\frac{1}{2N}(\partial_w h_{ij}-\bar\nabla_iN_j-\bar\nabla_jN_i)
\end{align}
and
\begin{align}
K_\varphi&=-\frac{1}{2N}\left(\partial_w-\mathcal L_{N^k}\right)\varphi\nonumber\\
&=-\frac{1}{2N}\left(\partial_w\varphi-N^k\partial_k\varphi\right)\ ,
\end{align}\label{defExtrinsicCurvatures}%
\end{subequations}
where $\mathcal L_{N^k}$ denotes the Lie derivative along the shift $N^k$.
As a result of this procedure we find:

\begin{widetext}
\begin{align}
N^{-1}\mathcal L&=\bar R+\epsilon\, \delta^{i_1i_2}_{j_1j_2}K^{j_1}_{i_1}K^{j_2}_{i_2}-2\bar\nabla^k\varphi\bar\nabla_k\varphi-8\epsilon K_\varphi^2\nonumber\\*
&-2\alpha\epsilon\,\delta^{i_1i_2i_3}_{j_1j_2j_3}\left[ \left(\epsilon\bar R^{j_1j_2}_{\ \ \ \ i_1i_2}+2K^{j_1}_{i_1}K^{j_2}_{i_2}\right)\bar\nabla^{j_3}\bar\nabla_{i_3}f(\varphi)+2f'(\varphi)K_\varphi K^{j_1}_{i_1}\left(\bar R^{j_2j_3}_{\ \ \ \ i_2i_3}-\frac{2\epsilon}{3}K^{j_2}_{i_2}K^{j_3}_{i_3}\right)\right]\nonumber\\
&+\alpha f(\varphi)\, \delta^{i_1i_2i_3i_4}_{j_1j_2j_3j_4}\left[\frac{1}{4}\bar R^{j_1j_2}_{\ \ \ \ i_1i_2}\bar R^{j_3j_4}_{\ \ \ \ i_3i_4}+\epsilon K^{j_1}_{i_1}K^{j_2}_{i_2}\left( \bar R^{j_3j_4}_{\ \ \ \ i_3i_4}-\frac{\epsilon}{3}K^{j_3}_{i_3}K^{j_4}_{i_4}\right)  \right]\ .\label{lagrangianADM}
\end{align}
\end{widetext}

As we shall check in Sec.~\ref{section3+1eom} below, the Euler-Lagrange equations derived from this Lagrangian return the $d+1$ decomposition of the equations $E^\mu_\nu=0$ and $E_\varphi=0$ [cf. Eqs.~(\ref{einsteinFieldEqn}) and (\ref{eomKGcovariante})] in the ADM variables (\ref{admCoordinates}).

For now, we rather focus on the associated Hamiltonian, defined as
\begin{equation}
\!\!\!H=\frac{1}{16\pi}\int_{\Sigma_w} \!\!\!\!\!d^dx \sqrt{|h|}\left(\pi_{ij}\,\partial_w h^{ij}+\pi_\varphi\,\partial_w\varphi-\mathcal L\right) .
\end{equation}
Since by definition we have
$\partial_w h^{ij}=-2N\!K^{ij}-\bar\nabla^iN^j-\bar\nabla^jN^i$ and $\partial_w\varphi=-2N\!K_\varphi+N^k\partial_k\varphi$,
an elementary calculation yields
\begin{equation}
H=\frac{1}{16\pi}\int_{\Sigma_w}\!\!\!\! d^dx\sqrt{|h|}\,( N\mathcal C+N^i\mathcal C_i)\ ,\label{ADMhamiltonian}
\end{equation}
where $\mathcal C$ and $\mathcal C_i$ are the Hamiltonian and momentum  constraints:
\begin{widetext}
\begin{subequations}
\begin{align}
\mathcal C &=-\bar R+\epsilon \, \delta^{i_1i_2}_{j_1j_2}K^{j_1}_{i_1}K^{j_2}_{i_2}+2\bar\nabla^k\varphi\bar\nabla_k\varphi-8\epsilon K_\varphi^2\nonumber\\
           &+2\alpha\,\delta^{i_1i_2i_3}_{j_1j_2j_3}\left[\left(\bar R^{j_1j_2}_{\ \ \ \ i_1i_2}-2\epsilon K^{j_1}_{i_1} K^{j_2}_{i_2}\right)\left(\bar\nabla^{j_3}\bar\nabla_{i_3}f(\varphi)-2\epsilon f'(\varphi)K_\varphi K^{j_3}_{i_3}\right)
             \right]\nonumber\\
&+\alpha f(\varphi)\, \delta^{i_1i_2i_3i_4}_{j_1j_2j_3j_4}\left[-\frac{1}{4}\bar R^{j_1j_2}_{\ \ \ \ i_1i_2}\bar R^{j_3j_4}_{\ \ \ \ i_3i_4}+\epsilon K^{j_1}_{i_1}K^{j_2}_{i_2}\left(\bar R^{j_3j_4}_{\ \ \ \ i_3i_4}-\epsilon K^{j_3}_{i_3}K^{j_4}_{i_4}\right)  \right]\ ,\label{perpConstraint}\\
\mathcal C_i &=2\bar\nabla^j\pi_{ij} +\pi_\varphi\,\bar\nabla_i\varphi\ .
\end{align}\label{contraintes}%
\end{subequations}
\end{widetext}

The Lagrangian (\ref{lagrangianADM}) and Hamiltonian (\ref{ADMhamiltonian}) are new, and their simple derivation based on integrating the ESGB momenta is the second main technical result of this paper.
Note that $H$ reduces to the Einstein-Gauss-Bonnet Hamiltonian when $f(\varphi)$ is a constant \cite{Teitelboim_1987,guilleminot}, and to the ADM Hamiltonian when $\alpha=0$ and $\varphi$ is a constant \cite{Arnowitt:1962hi}. Note also that the last lines of Eqs.~(\ref{lagrangianADM}) and (\ref{perpConstraint}) vanish identically when $D\leqslant 4$ (i.e. $d\leqslant 3$).

Since $H$ can only depend on the fields  and their conjugate momenta, the quantities $K^{ij}$ and $K_\varphi$ appearing in $\mathcal C$ above must be thought of as functions of $\pi_{ij}$ and $\pi_\varphi$, found by inverting the system of Eqs.~(\ref{momentumMetric}) and (\ref{momentumPhi}). However when $\alpha\neq 0$, $\pi_{ij}$ and $\pi_\varphi$ are nonlinear functions of $K^{ij}$ and $K_\varphi$.
Solving Eq.~(\ref{momentumPhi}) for $K_\varphi$ and substituting the result back into Eq.~(\ref{momentumMetric}) yields a system of $d(d+1)/2$ polynomial equations of degree five for $d(d+1)/2$ unknowns (the independent components of $K^{ij}$). In the weak Gauss-Bonnet coupling limit ($|\alpha\bar R^{ij}_{\ \, kl}|\ll1$ and $|\alpha^{1/2}K^{ij}|\ll1$) the solution can be approximated as a Taylor series, but the exact solution for a generic coupling $\alpha$ is not known in closed form (moreover, cf. \cite{RuffiniPaolo,AbelNiels,GaloisEvariste} for the existence of solutions in radicals to algebraic equations of degree five).\\

More importantly, the inversion of this system could have several real roots.
Therefore the Hamiltonian of ESGB theories is generically \textit{multivalued}. The same feature was previously discovered by Teitelboim and Zanelli in the case of Lovelock gravity in $D\geqslant 5$~\cite{Teitelboim_1987}, and it has two important implications:
\begin{itemize}
\item[(i)] at the classical level, the phase-space evolution of a system with initial data $(N,N^i;h^{ij},\varphi\,;\pi_{ij},\pi_\varphi)_{w_0}$, obtained by integrating Hamilton's equations,
can be \textit{unpredictable}, since the choice between different ``branches'' of the Hamiltonian is a priori arbitrary;
\item[(ii)] a multivalued Hamiltonian has serious shortcomings when attempting to canonically quantize the theory: see e.g.~\cite{Henneaux:1987zz,Shapere:2012yf,Zhao:2012dp,Avraham:2014twa,Ruz:2014ida} for simple toy models. The generalization of these toy models to ESGB theories is an interesting topic for future work.
\end{itemize}

Had ESGB theories been restricted to their weak Gauss-Bonnet coupling limit, points (i) and (ii) above would have been overlooked.

\section{The $d+1$ field equations of ESGB gravity\label{section3+1eom}}

In Sec.~\ref{sectionDirichlet} we built an action $I_{\rm ESGB}$ [Eq.~(\ref{actionMyers})] whose variation yields the covariant ESGB field equations $E^\mu_\nu=0$ and $E_\varphi=0$ when we impose Dirichlet boundary conditions. In Sec.~\ref{sectionADMformalism} we performed a $d+1$ decomposition of $I_{\rm ESGB}$ of the form (\ref{actionADM}),
where $\mathcal L=\mathcal L[N,N^i,h^{ij},\varphi]$ generalizes the ADM Lagrangian of general relativity to ESGB gravity, and is given in Eq.~(\ref{lagrangianADM}).

In this section we derive the equations of motion associated to the Euler-Lagrange variation of the action (\ref{actionADM}).
As we shall see, these equations of motion are the same as the $d+1$ decomposition of the covariant field equations $E^\mu_\nu=0$ and $E_\varphi=0$ using the ADM metric variables (\ref{admCoordinates}), as they should.

The action (\ref{actionADM}) does not depend on the normal derivatives $\partial_wN$ and $\partial_wN^i$. Therefore $N$ and $N^i$ are Lagrange multipliers, and the variation of (\ref{actionADM}) with respect to $\delta N$ and $\delta N^i$ yields the following $D$ constraints:
\begin{subequations}
\begin{align}
\mathcal C&=0\ ,\\
\mathcal C_i&=0\ ,
\end{align}\label{eomContraintes}%
\end{subequations}
where $\mathcal C$ and $\mathcal C_i$ are identical to the Hamiltonian and momentum constraints found in Eqs.~(\ref{contraintes}), which 
depend on the fields and on their first-order $w$-derivatives. In gaussian coordinates (such that $N^i=0$), they 
are equivalent to the set of Einstein equations $E^w_w=0$ and $E^w_i=0$: indeed, a short calculation using the Gauss-Codazzi-Mainardi identities (\ref{gaussCodazziMainardi}) gives
\begin{subequations}
\begin{align}
  E^w_w&=\mathcal C/2\,,\\
  E^w_i&=\mathcal C_i/2N\,.
\end{align}
 \label{EwwEwi}
\end{subequations}

The variation of the action (\ref{actionADM}) with respect to $h^{ij}$ and $\varphi$ yields the system of $\frac{1}{2}d(d+1)+1$ coupled equations
\begin{subequations}
\begin{align}
\frac{\partial\pi_{ij}}{\partial K_{ab}}\,\mathcal A_{ab}+\frac{\partial\pi_{ij}}{\partial K_\varphi}\, \mathcal A_\varphi&=\mathcal F_{ij}\ ,\\
\frac{\partial\pi_\varphi}{\partial K_{ab}}\,\mathcal A_{ab}+\frac{\partial\pi_\varphi}{\partial K_\varphi}\, \mathcal A_\varphi&=\mathcal F_\varphi\ ,
\end{align}\label{dynamicalEOM}%
\end{subequations}
where
\begin{subequations}
\begin{align}
 \hspace*{-0.32cm}\mathcal A_{ab} &=\frac{1}{N}(\partial_w-\mathcal L_{N^k})K_{ab} -K_{ak} K^k_b+\frac{\epsilon}{N}\bar\nabla_a\bar\nabla_b N ,\\
 \hspace*{-0.32cm}\mathcal A_\varphi &=\frac{1}{N}(\partial_w-\mathcal L_{N^k})K_\varphi \ .
\end{align}
\end{subequations}
The quantities $K_{ij}$ and $K_\varphi$ were defined in Eq.~(\ref{defExtrinsicCurvatures}), and furthermore
\begin{subequations}
\begin{align}
\hspace*{-0.3cm}\frac{\partial\pi^i_j}{\partial K^a_b}&=-\epsilon\delta^{ib}_{ja}+4\alpha\epsilon\,\delta^{ib\,i_1}_{ja\,j_1}\left(\bar\nabla^{j_1}\bar\nabla_{i_1}f(\varphi)-2\epsilon f'(\varphi)K_\varphi K^{j_1}_{i_1}\right)\nonumber\\*
\hspace*{-0.3cm}&-\alpha\epsilon f(\varphi) \delta^{ib\,i_1i_2}_{ja\,j_1j_2}\left(\bar R^{j_1j_2}_{\ \ \ \ i_1i_2}-2\epsilon K^{j_1}_{i_1} K^{j_2}_{i_2}\right),\label{dPidK1}\\*
\hspace*{-0.3cm}\frac{\partial\pi^i_j}{\partial K_\varphi} &=2\alpha\epsilon f'(\varphi)\delta^{i\,i_1i_2}_{j\,j_1j_2}\left(\bar R^{j_1j_2}_{\ \ \ \ i_1i_2}-2\epsilon K^{j_1}_{i_1} K^{j_2}_{i_2}\right),\\*
\hspace*{-0.3cm}\frac{\partial\pi_\varphi}{\partial K_\varphi}&=8\epsilon
\end{align}\label{dPidK}%
\end{subequations}
satisfy the integrability identities~(\ref{commutationMomenta}). The lengthy expressions for $\mathcal F_{ij}$ and $\mathcal F_\varphi$ will be given later for clarity.

Written as such, the structure of the dynamical equations of motion (\ref{dynamicalEOM}) is transparent.
They are {\em quasilinear} when $\alpha\neq 0$: that is, the coefficients of the ``accelerations'' $\mathcal A_{ab}$ and $\mathcal A_\varphi$ become functions of the fields $h^{ij}$ and $\varphi$ and of their first $w$-derivatives [cf. Eqs.~(\ref{dPidK})]. Let us set $n=d(d+1)/2+1$ and introduce the $n\times n$ matrix $\mathcal J$ with elements
\begin{align}
\mathcal J^I_{\ J}&=\begin{pmatrix}
\frac{\partial\pi^{ij}}{\partial K^{ab}} & \frac{\partial\pi^{ij}}{\partial K_\varphi}  \\
\frac{\partial\pi_\varphi}{\partial K^{ab}} & \frac{\partial\pi_\varphi}{\partial K_\varphi}
\end{pmatrix}\ ,\label{eq:jacobian}
\end{align}
where a capital index $I$ denotes either a pair of ordered indices $i\leqslant j$ or $\varphi$.
Inverting and evolving the system (\ref{dynamicalEOM}) on a constant-$w$ surface $\Sigma_w$ necessitates that the determinant of $\mathcal J$, i.e.
\begin{align}
\det\mathcal J=\frac{1}{n!}\,\delta^{I_1\cdots I_n}_{J_1\cdots J_n}\mathcal{J}^{J_1}_{\ I_1}\cdots \mathcal{J}^{J_n}_{\ I_n}\ ,\label{determinant}
\end{align}
\begin{comment}
[as in, e.g.,
\begin{subequations}
\begin{align}
\delta^{I}_{J}\mathcal{J}^{J}_{\ I}&={\rm Tr}\mathcal J=\delta^{i}_{a}\delta^{j}_{b}\frac{\partial\pi^{ab}}{\partial K^{ij}}+\frac{\partial\pi_\varphi}{\partial K_\varphi}\nonumber\\
&=8\epsilon+\epsilon d(d-1)\nonumber\\
&-4\alpha\epsilon (d-1)(d-2)\left(\bar\square f-2\epsilon f' K_\varphi K\right)\nonumber\\
&+2\alpha\epsilon f(\varphi) (d-2)(d-3)\nonumber\\
&\quad\times\left(\bar R-\epsilon (K^2-K^{ij}K_{ij})\right)\\
\delta^{I_1I_2}_{J_1J_2}\mathcal J^{J_1}_{\ I_1}\mathcal J^{J_2}_{\ I_2}&=({\rm Tr}\mathcal J)^2-{\rm Tr}\mathcal J^2\nonumber\\
&=({\rm Tr}\mathcal J)^2-\frac{\partial\pi^{ab}}{\partial K^{ij}}\frac{\partial\pi^{ij}}{\partial K^{ab}}\nonumber\\
&-2\frac{\partial\pi_\varphi}{\partial K^{ij}}\frac{\partial\pi^{ij}}{\partial K_\varphi}-\left(\!\frac{\partial\pi_\varphi}{\partial K_\varphi}\!\right)^{\!2} ,\end{align}
\end{subequations}
and so on]
\end{comment}
 \textit{be nonzero on $\Sigma_w$}.
Conversely, if there exists a location $x^i$ on the surface $\Sigma_w$ at which $\rm{det}\,\mathcal J=0$ [being understood that on-shell, the constraints (\ref{eomContraintes}) are satisfied], the dynamical equations of motion break down, and their predictability is lost. From Eqs.~(\ref{dPidK}), we see that this might happen in the nonperturbative Gauss-Bonnet coupling case, i.e. whenever $|\alpha\bar R^{ij}_{\ \, kl}|\gtrsim 1$ or $|\alpha^{1/2}K^{ij}|\gtrsim 1$.

In general, the explicit expression of $\det\mathcal J$ is cumbersome. When $D=4$ (i.e. $d=3$), for example, Eq.~(\ref{determinant}) is the determinant of a $7\times 7$ matrix. However, from the integrability equations (\ref{commutationMomenta}) it follows that $\mathcal J$ is a symmetric matrix,
\begin{align}
\mathcal J^I_{\ J}=\frac{\partial \pi^I}{\partial K^J}=\frac{\partial \pi_J}{\partial K_I}=\mathcal J^{\ I}_{J}\ ,
\end{align}
and it is hence diagonalizable. The specialization of our results to simpler isometric ``minisuperspaces'' is left for future work.

In $D\geq 5$, the quasilinearity of the Lovelock field equations and its consequences for the Cauchy problem and structure of characteristics was studied by Choquet-Bruhat in Refs.~\cite{ChoquetBruhat:acad,ChoquetBruhat:1988dw, 9780444860170}. As shown there, a quantity such as (\ref{determinant}) is a scalar on the $D$-dimensional spacetime $\mathcal M$: it only depends on the geometry of the foliation $\Sigma_w$ through invariant contractions of its extrinsic and intrinsic curvatures. These quantities can be written in arbitrary coordinates by the substitutions \cite{Wald:1984rg}: %
\begin{subequations}
\begin{align}
\hspace*{-0.1cm} K_{ij} &\to K_{\mu\nu}=\gamma_\mu^\lambda\nabla_\lambda n_\nu\ ,\\
\hspace*{-0.1cm}K_\varphi &\to K_\varphi=-\frac{1}{2}n^\lambda\nabla_\lambda\varphi\ ,\\
\hspace*{-0.1cm}\bar\nabla^i\bar\nabla_{\!j} f&\to \bar\nabla^\mu\bar\nabla_{\!\nu} f=\gamma^\mu_\alpha\gamma^\beta_\nu\nabla^\alpha\nabla_\beta f+2\epsilon f' \!K_\varphi K^\mu_\nu\, ,\\
\hspace*{-0.1cm}\bar R^{ij}_{\ \,kl} &\to \bar R^{\mu\nu}_{\ \ \rho\sigma}=\gamma^\mu_\alpha \gamma^\nu_\beta \gamma^\gamma_\rho \gamma^\delta_\sigma R^{\alpha\beta}_{\ \ \,\gamma\delta}\nonumber\\
\hspace*{-0.1cm}&\qquad\qquad\ +\epsilon\left(K^\mu_\rho K^\nu_{\sigma}-K^\mu_\sigma K^\nu_{\rho}\right)\, ,
\end{align}
\end{subequations}
where $n^\mu$ is a unit normal vector to $\Sigma_w$ such that $n^2=\epsilon$, given by $n^\mu=(\frac{1}{N},-\frac{N^i}{N})$ in ADM metric variables, and
 $\gamma^\mu_\nu=\delta^\mu_\nu-\epsilon n^\mu n_\nu$ is the projector on $\Sigma_w$.

For further developments on the problem of wave propagation in Lovelock gravity, see Refs.~\cite{Gibbons:1986uv,Tomimatsu:1987xy}; for an illustration of quasilinearity and its consequences in cosmology, see Refs.~\cite{Deruelle:1989fj,Deruelle:2003ck}. We also note that the $d+1$ decomposition of the Einstein-Gauss-Bonnet field equations was presented in Ref.~\cite{Torii:2008ru}.

In the present (ESGB) case, the right-hand sides of the dynamical field equations (\ref{dynamicalEOM}) read:
\begin{widetext}
\begin{subequations}
\begin{align}
\mathcal F^i_j &=-2\bar\nabla^i\varphi\bar\nabla_j\varphi+\delta^i_j\left(4\epsilon K_\varphi^2+\bar\nabla^k\varphi\bar\nabla_k\varphi\right)\nonumber\\
&+\delta^{i\,i_1i_2}_{j\,j_1j_2}\Bigg[\left(\bar R^{j_1j_2}_{\ \ \ \ i_1i_2}-2\epsilon K^{j_1}_{i_1}K^{j_2}_{i_2}\right)\left[-\frac{1}{4}+\alpha\left(4\epsilon f''(\varphi)K_\varphi^2+\frac{1}{N}\bar\nabla^k N\bar\nabla_k f\right)\right]\nonumber\\
&\qquad\qquad-4\alpha\epsilon\Big(K^k_{i_1}\bar\nabla_k f+2\bar\nabla_{i_1}(f'K_\varphi)\Big)\bar\nabla^{j_1}K^{j_2}_{i_2}-4\alpha\epsilon\Big(K_k^{j_1}\bar\nabla^k f+2\bar\nabla^{j_1}(f'K_\varphi)\Big)\bar\nabla_{i_1}K^{j_2}_{i_2}\Bigg]\nonumber\\
&+\alpha\,\delta^{i\,i_1i_2i_3}_{j\,j_1j_2j_3}\Bigg[\left(\bar R^{j_1j_2}_{\ \ \ \ i_1i_2}-2\epsilon K^{j_1}_{i_1}K^{j_2}_{i_2}\right)\left(\bar\nabla^{j_3}\bar\nabla_{i_3}f-2\epsilon f'(\varphi)K_\varphi K^{j_3}_{i_3}\right)+4\epsilon f(\varphi)(\bar\nabla^{j_1}K^{j_2}_{i_2})(\bar\nabla_{i_1}K^{j_3}_{i_3})\Bigg]\nonumber\\
&-\alpha f(\varphi)\, \delta^{i\,i_1i_2i_3i_4}_{j\,j_1j_2j_3j_4}\Bigg[\frac{1}{8}\bar R^{j_1j_2}_{\ \ \ \ i_1i_2}\bar R^{j_3j_4}_{\ \ \ \ i_3i_4}
-\frac{\epsilon}{2}K^{j_1}_{i_1}K^{j_2}_{i_2}(\bar R^{j_3j_4}_{\ \ \ \ i_3i_4}-\epsilon K^{j_3}_{i_3}K^{j_4}_{i_4})\Bigg]\ ,\label{einsteinDynamical}
\end{align}
\begin{align}
\mathcal F_\varphi &= 4\left(\bar\nabla^k\bar\nabla_k\varphi-2\epsilon K K_\varphi+\frac{1}{N}\bar\nabla^kN\bar\nabla_k\varphi\right)-8\alpha\epsilon f'(\varphi)\,\delta^{i_1i_2i_3}_{j_1j_2j_3}(\bar\nabla^{j_1}K^{j_2}_{i_2})(\bar\nabla_{i_1}K^{j_3}_{i_3})\nonumber\\
&+\alpha f'(\varphi)\,\delta^{i_1i_2i_3i_4}_{j_1j_2j_3j_4}\!\left(\frac{1}{4}\bar R^{j_1j_2}_{\ \ \ \ i_1i_2}\bar R^{j_3j_4}_{\ \ \ \ i_3i_4}-\epsilon K^{j_1}_{i_1}K^{j_2}_{i_2}(\bar R^{j_3j_4}_{\ \ \ \ i_3i_4}-\epsilon K^{j_3}_{i_3}K^{j_4}_{i_4})\right)\ .\label{KleinGordonDynamical}
\end{align}
\end{subequations}
\end{widetext}
Equations~(\ref{dynamicalEOM}) are the same as the decomposition of the Einstein and Klein-Gordon equations $E^i_j=0$ and $E_\varphi=0$ in ADM metric variables (\ref{admCoordinates}) obtained using the Gauss-Codazzi-Mainardi identities.

The ``$d+1$'' field equations of ESGB gravity (\ref{eomContraintes}) and (\ref{dynamicalEOM}) are the third, and main, result of this paper. 
They reduce to the field equations of general relativity when $\alpha=0$ and $\varphi$ is a constant~\cite{Gourgoulhon:1453298}, and they significantly simplify when $D\leqslant 4$ (i.e. $d\leqslant 3$), since then the last lines of Eqs.~(\ref{dPidK1}) and (\ref{KleinGordonDynamical}) and the last two lines of Eq.~(\ref{einsteinDynamical}) vanish identically.
Our results complement and extend previous works in various ways:
\begin{itemize}
  \item[(i)] we take into account the ESGB contributions of the scalar field and its coupling to the Gauss-Bonnet scalar, hence providing an explicit example of a ``quasilinear'' theory with nontrivial dynamics in dimension $D=4$ (i.e. $d=3$);
  
 \item[(ii)] we give, for the first time, the complete $d+1$ decomposition of the ESGB field equations, which could serve as a starting point to develop numerical relativity in these theories, extending the work of Refs.~\cite{Witek:2018dmd,Okounkova:2019zep,Okounkova:2019zjf,Okounkova:2020rqw};
 
 \item[(iii)] in our notation it should be clear that $\det\mathcal J$ is the Jacobian of the change of variables $(K^{ij},K_\varphi)\!\to\!(\pi^{ij},\pi_\varphi)$. Therefore if $\det\mathcal J$ vanishes at any point $x^\mu=(w,x^i)$, not only the predictability of the dynamical equations of motion (\ref{dynamicalEOM}) is lost, but also the Hamiltonian is not defined, since the change of variables is noninvertible.
\end{itemize}

We conclude this section by establishing the constraint propagation equations of ESGB gravity, which can be found by a direct generalization of their general relativistic counterpart.
From the Bianchi identities we have that $\nabla_\mu G^\mu_\nu=\nabla_\mu P^\mu_{\ \,\nu\rho\sigma}=\nabla_\mu H^\mu_\nu=0$, as discussed below Eq.~(\ref{lanczosTensor}).
Therefore, taking the divergence of $E^\mu_\nu$ in Eq.~(\ref{einsteinFieldEqn}),  a simple calculation %
 yields
\begin{equation}
\nabla_\mu E^\mu_\nu=-\frac{1}{2}E_\varphi\nabla_\nu\varphi\ ,
\end{equation}
where $E_\varphi$ is given in Eq.~(\ref{eomKGcovariante}). Now we can choose gaussian coordinates (\ref{gaussianCoord}) for simplicity, and evaluate the $\nu=w$ and $\nu=i$ components of the identity above.
Using Eqs.~(\ref{EwwEwi}) we find
\begin{subequations}
\begin{align}
(\partial_w-\mathcal L_{N^k})\,\mathcal C&=-N\!K\mathcal C-2\epsilon \,\mathcal C_i\bar\nabla^iN-\epsilon N\bar\nabla^i\mathcal C_i\nonumber\\
&+2N\!K^i_jE^j_{\,i}+2NK_\varphi E_\varphi\ ,\\
(\partial_w-\mathcal L_{N^k})\,\mathcal C_i&=-\,\mathcal C_iN\!K+\mathcal C\bar\nabla_iN\nonumber\\
&-2\bar\nabla_j(NE^j_{\,i})-NE_\varphi\bar\nabla_i\varphi\ ,
\end{align}\label{constraintPropag}%
\end{subequations}
where we restored the shift $N^k$ for completeness.

If the dynamical equations of motion are satisfied, i.e. $E^i_j=E_\varphi=0$ [see below Eq.~(\ref{KleinGordonDynamical})], then the last lines of Eq.~(\ref{constraintPropag}) vanish.
Therefore, if the constraints are satisfied on a surface $\Sigma_{w_i}$ with $w=w_i$,
\begin{equation}
\mathcal C|_{w_i}=\mathcal C_i|_{w_i}=\partial_j\mathcal C_i|_{w_i}=0\ ,
\end{equation}
they are satisfied on every surface $\Sigma_w$.

\section{Conclusions}
\label{sec:conclusions}

In this paper we have presented the $d+1$ formulation of ESGB theories in dimension $D=d+1$ and for arbitrary (spacelike or timelike) slicings. Our main results are:
\begin{itemize}
  \item[1)] An extension of the actions found by Gibbons-Hawking-York \cite{Gibbons:1976ue,York:1972sj} and Myers \cite{Myers:1987yn} to ESGB gravity theories [Eq.~(\ref{actionMyers})];
  \item[2)] The ADM Lagrangian (\ref{lagrangianADM}) and Hamiltonian (\ref{ADMhamiltonian}), which we found in a simple manner by integrating the ESGB momenta;
  \item[3)] The $d+1$ decomposition of the ESGB field equations [Eq.~(\ref{dynamicalEOM})] and the corresponding constraint propagation equations [Eqs.~(\ref{constraintPropag})].
\end{itemize}

Our results should be useful to guide future developments of numerical relativity for ESGB gravity in the nonperturbative regime, and eventually to obtain gravitational waveforms for the whole inspiral, merger and ringdown of compact binary systems. This is important, because most analytical~\cite{Yagi:2011xp,Yagi:2015oca} and numerical~\cite{Witek:2018dmd,Okounkova:2019zep,Okounkova:2019zjf,Okounkova:2020rqw} studies of black-hole binaries in ESGB 
so far have relied on a small-$\alpha$ expansion. If and when numerical relativity waveforms become available, it will be interesting to compare them with post-Newtonian calculations valid in the inspiral phase~\cite{Julie:2019sab} and to guide developments of an effective-one-body model for ESGB gravity, similar in spirit to previous work in scalar-tensor~\cite{Julie:2017pkb,Julie:2017ucp} and Einstein-Maxwell-dilaton~\cite{Julie:2017rpw,Julie:2018lfp} theories of gravity.

Pretorius and Ripley~\cite{Ripley:2019hxt,Ripley:2019irj,Ripley:2019aqj} have recently studied spherically symmetric collapse in ESGB, finding evidence that there are open sets of initial data for which the character of the system of equations changes from hyperbolic to elliptic in a compact region of the spacetime.  It will be interesting to specialize our equations of motion to spherically symmetric spacetimes and further investigate this loss of hyperbolicity.  

We noted that the nonlinearity of the momenta plays a role in both the Hamiltonian's multivaluedness and the quasilinearity of the dynamical equations of motion. Moreover, when $\rm{det} \,\mathcal J$ defined in Eq.~(\ref{eq:jacobian}) vanishes at a point $x^i$ of a constant-$w$ foliation $\Sigma_w$, the predictability of the dynamical equations of motion (\ref{dynamicalEOM}) is lost and the Hamiltonian is not defined.
Since $\rm{det} \,\mathcal J$ must be evaluated on-shell,  it will be very useful to carefully study the role of the constraints on the values this determinant can take.  Furthermore, we conjecture that these issues (the existence of a multivalued Hamiltonian and a possible breakdown of the Lagrangian equations of motion) may be generic features of higher-order theories, such as Horndeski theories of gravity.

We also expect the multivaluedness of the Hamiltonian $H$ to be related to pathologies at the quantum level.  These considerations imply that ESGB gravity should only be considered as an effective low-energy field theory. Note however that ESGB gravity (as well as other higher-order theories, such as Lovelock and Horndeski theories) is, by construction, devoid of ghosts, which is indeed a requirement to build a quantum theory \cite{Woodard:2009ns}.  Our work suggests that the nonmultivaluedness of $H$ should be treated as another important selection criterion for modified theories of gravity, possibly as important as the absence of ghosts.

Finally, a possible extension of our results is to generalize the definition of the ADM mass to ESGB theories. This would be very useful to define global charges of BHs, which play a central role in their thermodynamics and in other applications, such as gauge-gravity dualities in arbitrary dimension. These issues will be addressed in future work.

\acknowledgments
We are grateful to Nathalie Deruelle for enlightening discussions and suggestions along the preparation of this work. We also thank \'Eric Gourgoulhon, Nelson Merino, Hector O. Silva, Thomas Sotiriou, Helvi Witek and Nicol\'as Yunes for discussions.
F.L.J. and E.B. are supported by NSF Grants No. PHY-1912550 and AST-1841358, NASA ATP Grants No. 17-ATP17-0225 and 19-ATP19-0051, and NSF-XSEDE Grant No. PHY-090003. The authors would like to acknowledge networking support by the GWverse COST Action CA16104, ``Black holes, gravitational waves and fundamental physics'' and from the Amaldi Research Center funded by the MIUR program ``Dipartimento di Eccellenza''~(CUP: B81I18001170001).

\bibliography{FLJbib}

\end{document}